\documentclass[pra,aps,showpacs,twocolumn,superscriptaddress,amsmath,amssymb]{revtex4}

\usepackage{graphicx,dcolumn,bm}
\newcommand {\startbild}{\begin{figure}}
\newcommand {\stopbild}{\end{figure}}

\newcommand {\staf}{\begin{equation}}
\newcommand {\stof}{\end{equation}}
\newcommand {\staffeld}{\begin{eqnarray}}
\newcommand {\stoffeld}{\end{eqnarray}}

\newcommand{\ket}[1]{|#1\rangle}
\newcommand{\bra}[1]{\langle #1|}
\renewcommand{\vec}[1]{{\bf #1}}
\begin{document}

\title{Creating path entanglement and violating Bell inequalities \\by independent photon sources}

\author{R. Wiegner}
\email{ralph.wiegner@physik.uni-erlangen.de}
\affiliation{Institut f\"ur Optik, Information und Photonik, Universit\"at Erlangen-N\"urnberg, Erlangen, Germany}
\homepage{http://www.ioip.mpg.de/jvz/}

\author{C. Thiel}
\affiliation{Institut f\"ur Optik, Information und Photonik, Universit\"at Erlangen-N\"urnberg, Erlangen, Germany}

\author{J. von Zanthier}
\affiliation{Institut f\"ur Optik, Information und Photonik, Universit\"at Erlangen-N\"urnberg, Erlangen, Germany}

\author{G. S. Agarwal}
\affiliation{Department of Physics, Oklahoma State University, Stillwater, OK, USA}

\date{\today}

\begin{abstract}

We demonstrate a novel approach of violating position dependent Bell inequalities by photons emitted via independent photon sources in free space. We trace this violation back to path entanglement created a posteriori by the selection of modes due to the process of detection.

\end{abstract}


\maketitle

\section{Introduction}

Path and polarization entangled photons play a key role in the investigations of basic aspects of quantum mechanics \cite{CITE1}. They are also fundamental for applications in quantum information processing \cite{CITE2}. Until now the predominant source of entangled photons have been  nonlinear crystals, employed for spontaneous parametric down conversion (SPDC). Via SPDC, from an initial pump photon, path and polarization entangled photons can be created and further processed, e.g., to engineer more complex multiphoton entangled states \cite{4,5,weinfurter1}. 

Apart from non-linear crystals, also uncorrelated single photon emitters have been proposed as a source to generate path and polarization entangled photonic states \cite{yurke,7,Dowling2002,takeuchi0,takeuchi1,11,agarwal2002,christoph2,james}. Hereby, either linear optical tools or filters can be employed to produce the entanglement \cite{yurke,7,Dowling2002,takeuchi0,takeuchi1}, but there are also proposals to generate entangled photons without inserting any element, i.e., in free space \cite{11,agarwal2002,christoph2,james}.

In this paper we investigate how photons can become path entangled in free space if they are emitted by initially uncorrelated single photon emitters. The path entanglement is demonstrated by use of the second order correlation function violating a position dependent Bell inequality \cite{bell}. It is shown that the entanglement originates out of an initially separable state with the help of mode selection induced by the process of detection.

The paper is organized as follows: In sections \ref{SUI} we introduce the physical setup investigated throughout this paper. In section \ref{bell1} we discuss a novel approach to violate Bell-type inequalities \cite{ch} using position variables. Section \ref{CEMS} introduces a quantum path description to demonstrate that out of a separable initial state, consisting of an infinity of potentially occupied modes, two entangled quantum paths are selected due to the process of detection. Moreover, we compare our method to SPDC. In section \ref{conclusion} we finally conclude.

\section{System under investigation}
\label{SUI}

\begin{figure}[h!]
\centering
\includegraphics[width=0.4 \textwidth, bb=0 0 578 327]{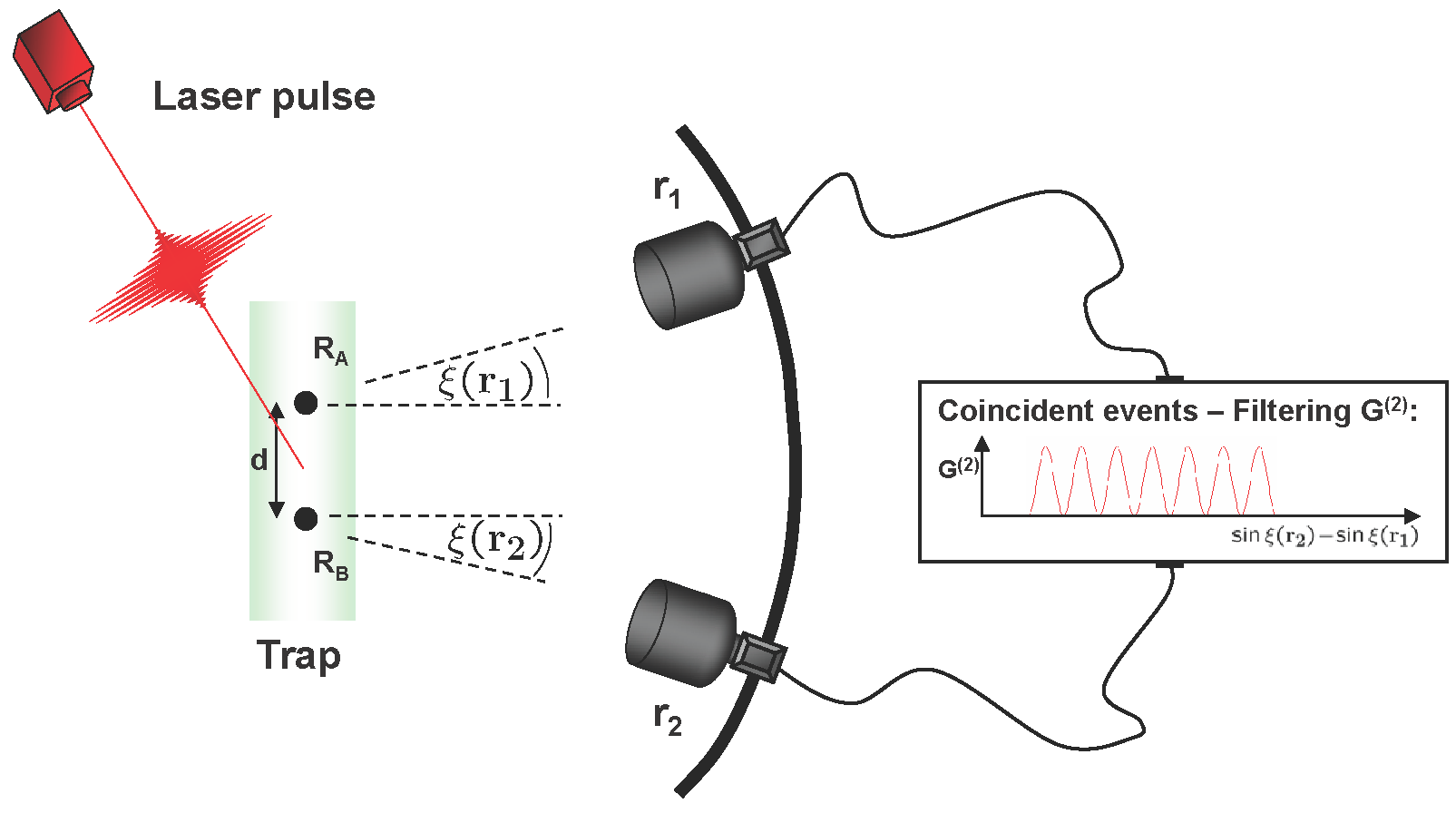}
\caption{\label{CQI} Two uncorrelated two-level atoms are localized at positions ${\bf R}_A$ and ${\bf R}_B$. Both atoms are initially fully excited and emit two photons via spontaneous decay. The photons are detected in the far field by two detectors at positions ${\bf r}_1$ and ${\bf r}_2$.}
\end{figure}

To demonstrate the appearance of entanglement among photons emitted by initially uncorrelated sources in its simplest form, we consider two two-level atoms with upper level $\ket{e}$ and ground state $\ket{g}$, localized at positions ${\bf R}_A$ and ${\bf R}_B$ with a separation $d >\!> \lambda$, so that the dipole-dipole interaction can be neglected (cf.~figure \ref{CQI}). The atoms are initially fully excited to the state $|e\,e\rangle\equiv|e\rangle_A\otimes|e\rangle_B$, e.g., by a single laser $\pi$ pulse. After the excitation each atom spontaneously emits a single photon. The two photons are assumed to be registered by two detectors at positions ${\bf r}_1$ and ${\bf r}_2$, where $|\vec{r}_i - \vec{R}_n| \gg d$ ($i = 1,2$, $n = A, B$), so that the far field condition is fulfilled.

To simplify further calculations we only consider coincident detections, i.e., simultaneous events at both detectors. However, we note that coincident detection is not a prerequisite for our scheme since the detection time does not influence the modulation nor the contrast of the correlation signal \cite{Agarwal74,Agarwal77}. Furthermore, we assume that each of the two detectors registers exactly one photon so that two photon absorption processes at one detector are excluded. This can be simply implemented via post-selection.

As we want to investigate the photonic correlations, we make use of the correlation functions introduced by Glauber \cite{glauber1}. The first and the second order correlation function of the electric field can be written as \cite{Agarwal74}

\begin{eqnarray}\label{tutg1}
&&G^{(1)}({\bf r}_1)= \langle \hat{{\bf E}}^{(+)}\!({\bf r}_1)\hat{{\bf E}}^{(-)}\!({\bf r}_1)\rangle\\
&&G^{(2)}({\bf r}_1,{\bf r}_2)=\nonumber\\
&&\langle \hat{{\bf E}}^{(+)}\!({\bf r}_1)\hat{{\bf E}}^{(+)}\!({\bf r}_2)\hat{{\bf E}}^{(-)}\!({\bf r}_2)\hat{{\bf E}}^{(-)}\!({\bf r}_1)\rangle.
\end{eqnarray}

\noindent Hereby, the electric field operator $\hat{E}^{(-)}({\bf r}_i)$ takes the form

\begin{eqnarray}
\hat{{\bf E}}^{(-)}({\bf r}_i)=\frac{E_0}{\sqrt{2}} \left(e^{-ik({\bf \hat{r}}_i\cdot{\bf R}_A)}\cdot{S}_A^-+e^{-ik({\bf \hat{r}}_i\cdot{\bf R}_B)}\cdot{S}_B^-\right),
\end{eqnarray}
where $k=\frac{2\pi}{\lambda}$ denotes the wave number of the transition $|e\rangle\rightarrow|g\rangle$, ${\bf \hat{r}}_i:=\frac{{\bf r}_i}{|{\bf r}_i|}$ is a unit vector in the direction of the $i$th detector, ${\bf R}_n$ specifies the position of the $n$th atom ($n = A, B$), the amplitude of the electric field is abbreviated by $E_0$ and ${S}_A^-$ and ${S}_B^-$ denote the lowering operator $\ket{g}\bra{e}$ of the transition $|e\rangle\rightarrow|g\rangle$ of atom $A$ and $B$, respectively~\cite{11}. 

As the atoms are initially fully excited to the state $|e\,e\rangle$ and taking only coincident detection into account, the $G^{(1)}$- and $G^{(2)}$-function can be calculated to \cite{Agarwal2004}

\begin{eqnarray}\label{G1calc}
G^{(1)}(\vec{r}_1)&=&\left|\,\hat{E}^{(-)}(\vec{r}_1)|e\rangle|e\rangle\,\right|^2 = E_0^2\\
\label{G2calc}
G^{(2)}(\vec{r}_1, \vec{r}_2)&=&\left|\,\hat{E}^{(-)}(\vec{r}_2)\hat{E}^{(-)}(\vec{r}_1)|e\rangle|e\rangle\,\right|^2\nonumber\\
&=&\frac{E_0^4}{2}\,(1 + \cdot\cos\left[\left(\vec{r}_2-\vec{r}_1\right)\cdot k \vec{d}\right])\;,
\end{eqnarray}
where $\vec{d} = {\bf R}_B-{\bf R}_A$ denotes the distance vector between the two atoms. After introducing the visibility ${\cal V}$ which, due to a limited temperature of the ions \cite{Wineland}, extended detector sizes \cite{Uwe}, or other experimental imperfections, takes on values ${\cal V} \leq 1$, we arrive at the following expression for $G^{(2)}(\vec{r}_1, \vec{r}_2)$:
\begin{eqnarray}\label{G2V}
G^{(2)}(\vec{r}_1, \vec{r}_2)&= &\frac{E_0^4}{2}\,(1 + {\cal V}\cdot\cos\left[\left(\vec{r}_2-\vec{r}_1\right)\cdot k \vec{d}\right])\;,
\end{eqnarray}
Interpreting the first and the second order correlation function from a probabilistic point of view, we obtain the identities

\staffeld
\label{QMPROB1}
P(\vec{r}_1) & = & \frac{\eta}{E_0^2} \, G^{(1)}(\vec{r}_1),\\
\label{QMPROB2}
\label{QMPROB3}
P_{12}(\vec{r}_1,\vec{r}_2) & = & \frac{\eta^2}{E_0^4} \, G^{(2)}(\vec{r}_1, \vec{r}_2).\;
\stoffeld

\noindent Hereby, $P(\vec{r}_1)$ abbreviates the probability to detect a photon at position $\vec{r}_1$, $P_{12}(\vec{r}_1,\vec{r}_2)$ denotes the joint probability to detect a photon at position $\vec{r}_1$ and another photon at position $\vec{r}_2$ and $\eta$ incorporates experimental insufficiencies, e.g., the quantum efficiency of the detectors, and the angle subtended by the detectors. We note that the joint probability $P_{12}(\vec{r}_1,\vec{r}_2)$ can also be written in the form \cite{11}

\begin{eqnarray}
\label{CPB1}
P_{12}({\bf r}_1,{\bf r}_2)=P({\bf r}_2|{\bf r}_1) \cdot P({\bf r}_1),
\end{eqnarray}

\noindent where $P({\bf r}_2|{\bf r}_1)$ is the conditional probability to find a photon at position ${\bf r}_2$ if another one is detected at position ${\bf r}_1$. As the $G^{(1)}$-function is a constant (cf.~Eq.~(\ref{G1calc})), the second order correlation function can directly be linked to $P({\bf r}_2|{\bf r}_1)$, i.e., we have

\staf
\label{g2G2}
G^{(2)}({\bf r}_1,{\bf r}_2) \sim P({\bf r}_2|{\bf r}_1).
\stof

Note that in case of ${\cal V}=100\%$, due to the cosine modulation of the $G^{(2)}$-correlation function in~Eq.~(\ref{G2V}), there are positions ${\bf r}_2$ where no photon can be detected if another photon is measured at a particular position ${\bf r}_1$. With the relation given in Eq.~(\ref{g2G2}) we thus convey that a highly non-local, i.e., a highly non-classical behaviour is displayed in this case by the system.

To investigate this in more detail we will quantify in the next section the non-classicality exhibited by the second order correlation function. For this purpose we will make use of a set of Bell-type inequalities, the so-called CH74 inequalities \cite{ch}, and choose appropriate variables in order to prove that correlations among the photons are present in our system which violate the predictions of classical physics.

\section{A new approach to violate Bell inequalities with position variables}
\label{bell1}

In 1935 Einstein, Podolsky and Rosen published their well-known article as an argument for the necessity to introduce supplementary parameters $\lambda$ in order to complete quantum mechanics~\cite{EPR}. These so called hidden variables were advanced to avoid the non-local nature of the quantum mechanical description, i.e., to restore \emph{locality} in the theory. Locality assumes that the outcome of a measurement at a particular position is independent of the measurement at another position. In our setup this means that the probability of detecting a single photon at position $\vec{r}_1$ is independent of detecting another photon at position $\vec{r}_2$.\\

Many different forms of Bell inequalities can be derived, which allow to distinguish between a local realistic theory and quantum mechanics. In our setup the CH74 inequalities \cite{ch} seem the most appropriate. They read
 
\staf
\label{ch1}
-XY\leq x\,y-x\,y'+x'\,y+x'\,y'-x'\,Y-y\,X\leq0,
\stof

\noindent and hold for 

\staf
\label{ass1}
0\leq x,x'\leq X \wedge 0\leq y,y'\leq Y.
\stof


Until now there have been proposals which describe a position dependent violation of the CH74 inequalities with the help of polarization degrees of freedom (cf., e.g., \cite{ou}). These auxilliary degrees of freedom are needed in order to arrive at some constant probabilities which allow for a normalization of the inequalities (\ref{ch1})on experimental parameters \cite{ch,ou}. However, in our considered two-level system it is also possible to normalize these inequalities without taking into account auxilliary degrees of freedom. To this end we consider single mode fibers between the atoms and the detectors. The joint probability of detecting two photons scattered by two atoms with one fiber leading from atom $A$ to detector $1$ and one fiber leading from atom $B$ to detector $2$ calculates to

\staf
\label{star1}
P_{12}(\star,\star)  =  \frac{\eta^2}{E_0^4} \cdot E_0^4
\stof 

where a $\star$ stands for a joint probability which is independent on the corresponding detector position. If we relate the probability $P_{12}(\star,\star)$ to the parameters $X$ and $Y$ of the inequalities (\ref{ch1}), we arrive at

\staffeld\label{mimic}
- P_{12}(\star,\star)\leq P_{12}(\vec{r}_1,\vec{r}_2)- P_{12}(\vec{r}_1,\vec{r}_2')+ P_{12}(\vec{r}_1',\vec{r}_2)\nonumber\\
+ P_{12}(\vec{r}_1',\vec{r}_2')- P_{12}(\vec{r}_1',\star)- P_{12}(\star,\vec{r}_2) \leq 0 ,\;
\stoffeld
where $P_{12}(\vec{r}_1',\star) = P_{12}(\star,\vec{r}_2) = P_{12}(\star,\star) = \eta^2$. By inserting the corresponding joint probabilities and constant probabilities into Eq.~(\ref{mimic}) we obtain for the upper bound of the CH74 inequalities

\begin{eqnarray}\label{mimicUB}
&&\frac{1}{2}{\cal V}\left(\cos[\frac{\pi}{4}] - \cos[\frac{3}{4}\pi] + \cos[\frac{\pi}{4}] + \cos[\frac{\pi}{4}]  \right) - 1\leq 0 \nonumber\\
&& \hspace{2cm}\rightarrow {\cal V} \cdot\sqrt{2} -1 \leq 0, 
\end{eqnarray}

where the well known Bell angles ($\frac{\pi}{4}, \frac{3\pi}{4}, \frac{\pi}{4}, \frac{\pi}{4}$) (cf., e.g., \cite{bertlmann}) have been chosen for our position variables. A maximal violation \cite{maxvio} of the upper bound of the CH74 inequalities is thus obtained in case of a visibility ${\cal V} > \frac{1}{\sqrt{2}} \approx 71\%$, what proves the existence of non-classical correlations among the emitted photons, i.e., a photon path entanglement in our system. This result is in agreement with previous investigations, however without the need for auxilliary polarization degrees of freedom. We note that restricting our proposal to the upper bound of the CH74 inequalities is not a major drawback as the lower bound is not as stringent as the upper bound \cite{C78}. In the following section we will discuss how it is possible that two photons, spontaneously emitted by two initially  entirely uncorrelated atoms, can give rise to a second order correlation signal displaying highly non-classical correlations among the registered photons.

\section{Creation of path entanglement by mode selection}
\label{CEMS}

\begin{figure}[h!]
\centering
\includegraphics[width=0.17 \textwidth, bb=0 0 267 488]{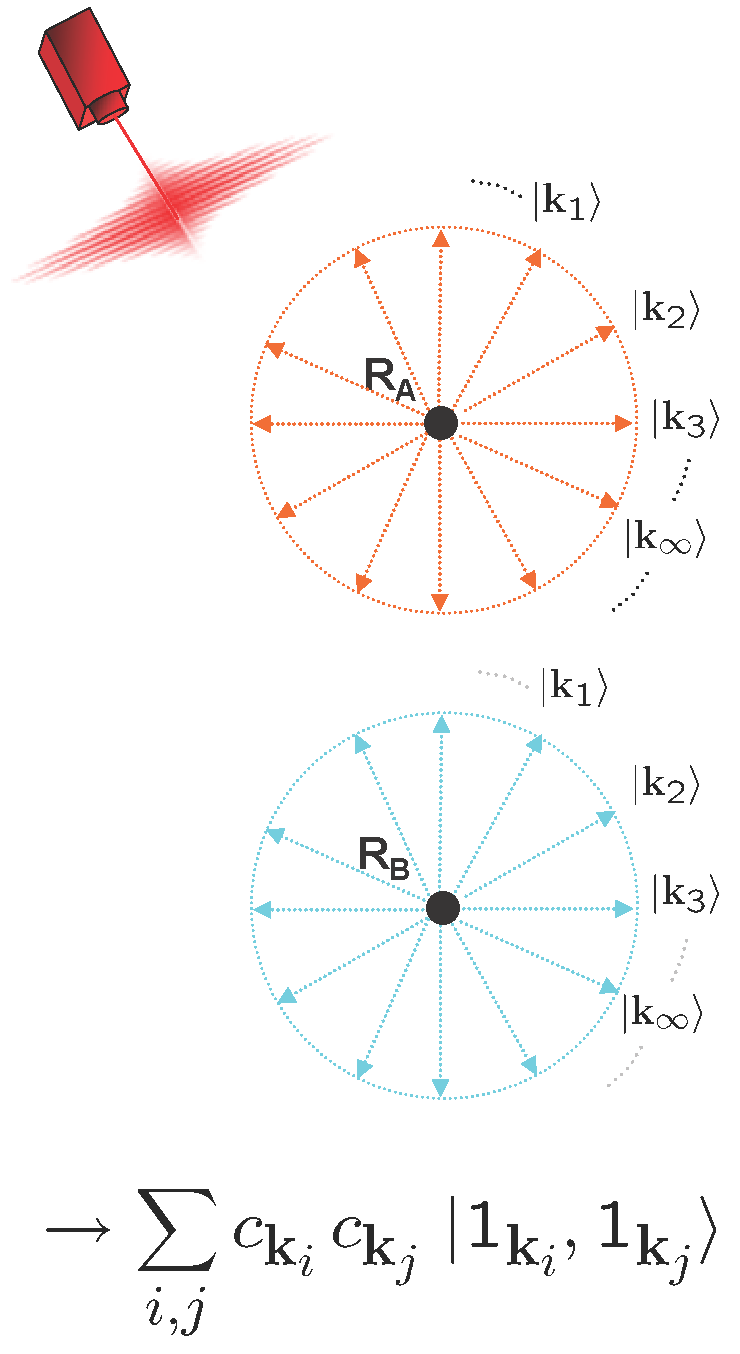}
\caption{\label{fullspace} Each of the two initially fully excited atoms located at position ${\bf R}_{A}$ and ${\bf R}_{B}$ scatters a photon. After a time $t$ assumed to be much greater than the decay time $\tau$ of the atoms the photons are in the state $\sum_{i,j} c_{\vec{k}_i}\,c_{\vec{k}_j}\;\ket{1_{\vec{k}_i},1_{\vec{k}_j}}$ (cf.~text for details).}  
\end{figure}

In order to investigate this question let us start by formalizing the process of spontaneous decay of a single atom. If the atom located at position ${\bf R}_A$ is initially excited and, after scattering a photon by spontaneous decay, returns to the ground state $\ket{g}$, an infinity of photonic modes in a mode space $K$ is potentially occupied. The same is true for the other atom located at position ${\bf R}_B$: if this atom scatters a photon and returns to the ground state $\ket{g}$, also an infinity of photonic modes in the mode space $K$ is potentially occupied (cf. figure \ref{fullspace}).

After a time $t$ - assumed to be much greater than the decay time $\tau$ - both atoms will be found in the ground states $\ket{g\;g}$ whereby each spontaneously scattered photon populates potentially the entire accessible mode space $K$. For the two photons we thus obtain the following photonic state 

\staf
\label{varphi3'}
\ket{\varphi}_{1} = \sum_{i,j} c_{\vec{k}_i}\,c_{\vec{k}_j}\;\ket{1_{\vec{k}_i},1_{\vec{k}_j}}\;,
\stof
where the coefficients $c_{\vec{k}_j}$ can be obtained from Wigner Weisskopf theory \cite{scully} and $\ket{1_{\vec{k}_i},1_{\vec{k}_j}}$ is given by

\begin{align}
\label{OC}
&\ket{1_{\vec{k}_i},1_{\vec{k}_j}} \equiv \ket{0_{\vec{k}_1}, 0_{\vec{k}_2},...,1_{\vec{k}_i},0_{\vec{k}_{i+1}},..., 1_{\vec{k}_j}, 0_{\vec{k}_{j+1}},..., 0_{\vec{k}_m}}\nonumber \\ 
&m \rightarrow \infty,
\end{align}

\noindent abbreviating that the modes $\vec{k}_i$ and $\vec{k}_j$ are occupied with one photon and all other modes in $K$ are empty.

The state $\ket{\varphi}_{1}$ describes the photonic degrees of freedom incorporating all possibilities of two arbitrary modes being populated in the mode space $K$ and not yet being detected. Clearly the state $\ket{\varphi}_{1}$ is separable. It thus appears not intuitive to obtain a second order correlation signal violating the Bell inequalities as in  Eq.~(\ref{G2calc}).

\begin{figure}[h!]
\centering
\includegraphics[width=0.4\textwidth, bb=0 0 1277 638]{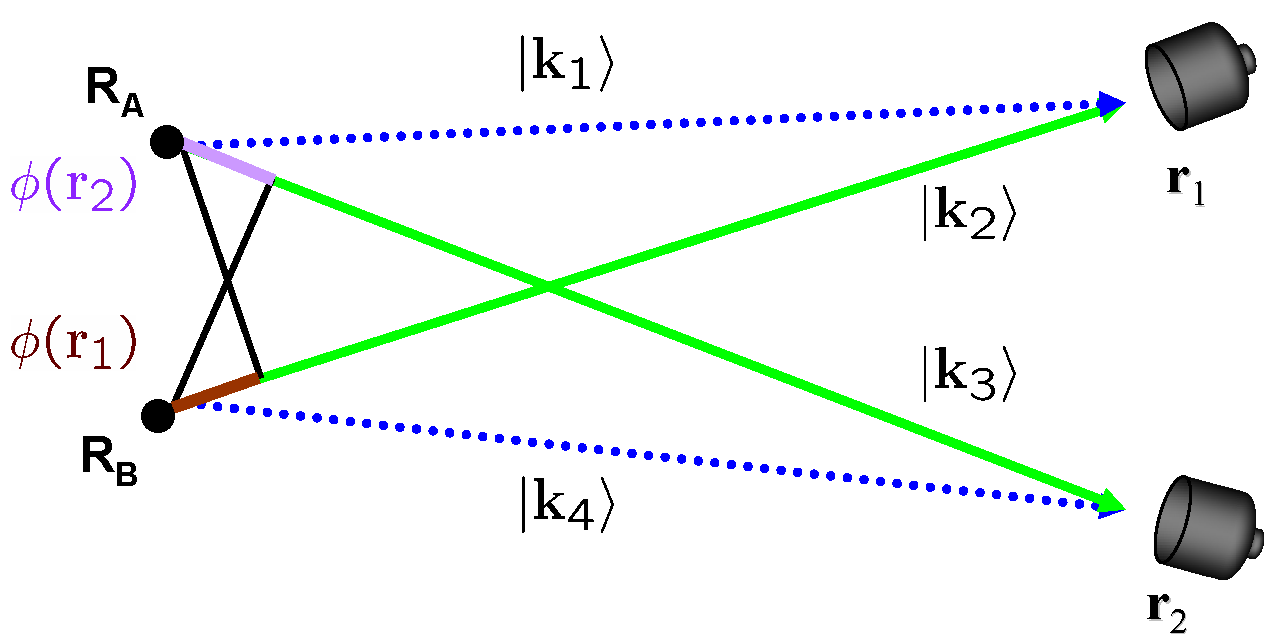}
\caption{\label{reducedspace} If two detectors are located in the far field and assuming that each detector will register one photon, the photonic mode space is reduced to the four modes $\vec{k}_1$ to $\vec{k}_4$. There is a relative phase $e^{i\phi({\bf r}_1)}$ ($e^{i\phi({\bf r}_2)}$) between the modes $\vec{k}_1$ and $\vec{k}_2$ ($\vec{k}_3$ and $\vec{k}_4$) depending on the detector position ${\bf r}_{1}$ (${\bf r}_{2}$).}
\end{figure}

However, this is the situation if the two photons have not yet been detected and therefore occupy potentially all modes. If on the other hand we require via post-selection that for a successful measurement each of the two detectors register exactly one photon we reduce the photonic mode space considerably and obtain a state of the following form before the detection of the photons (cf. figure \ref{reducedspace})

\staf
\label{varphi1'}
\ket{\varphi}_{1'} = \ket{1_{\vec{k}_1}\;0_{\vec{k}_2}\;0_{\vec{k}_3}\;1_{\vec{k}_4}} + \ket{0_{\vec{k}_1}\;1_{\vec{k}_2}\;1_{\vec{k}_3}\;0_{\vec{k}_4}}
\stof
with $\ket{1_{\vec{k}_i}}$ ($\ket{0_{\vec{k}_i}}$) abbreviating that the mode $\vec{k}_i$ is occupied with 1 (0) photons ($i = 1, \ldots, 4$). For simplicity we will suppress the mode indices $\vec{k}_i$ in the following and write the state $\ket{\varphi}_{1'}$ in the shortened form $\ket{\varphi}_{1'} = \ket{1\;0\;0\;1}+ \ket{0\;1\;1\;0}$.

As initially stated, due to the far field detection scheme, no precise which-way information can be obtained for the photons so that the two emitted photons have exactly two possibilities to propagate from the two atoms towards the two detectors: either via the blue (dotted lines) quantum path $\ket{1_{\vec{k}_1}\;1_{\vec{k}_4}}$ or the green (solid lines) quantum path $\ket{1_{\vec{k}_2}\;1_{\vec{k}_3}}$ as displayed in figure \ref{reducedspace}. 

In the first two-photon quantum path one photon populates mode $\vec{k}_1$ \emph{and} the other photon populates mode $\vec{k}_4$ (the blue lines). The other two-photon quantum path represents the scenario that one photon populates mode $\vec{k}_2$ \emph{and} the other photon populates mode $\vec{k}_3$ (the green lines). 

Without loss of generality we now assume that the first detector located at position ${\bf r}_1$ registers a photon. This process can be described by the action of a detector operator $\hat{D}({\bf r}_1)$ on the state $\ket{\varphi}_{1'}$, where $\hat{D}({\bf r}_1)$  is given by

\staf
\hat{D}({\bf r}_1) = \ket{0\;0\;0\;1}\bra{1\;0\;0\;1} + e^{i\phi({\bf r}_1)} \ket{0\;0\;1\;0}\bra{0\;1\;1\;0}\;.
\stof
The detector measures and thus destroys and removes one photon from the occupied photonic mode $\vec{k}_1$ or - with a relative phase $e^{i\phi({\bf r}_1)}$ depending on the detector position ${\bf r}_1$ - from the occupied photonic mode $\vec{k}_2$ (cf. figure \ref{reducedspace}). Hereby, the relative phase $\phi({\bf r}_i)$ is given by $\phi({\bf r}_i) = dk\,\sin(\xi(\vec{r}_i))$ (cf.~figure \ref{CQI}).

The state after the detection of the first photon thus can be calculated to

\staffeld
\label{varphi4'}
\ket{\varphi}_{2} = \hat{D}({\bf r}_1)\,\ket{\varphi}_{1'} = \ket{0\;0\;0\;1} + e^{i\phi(\vec{r}_1)} \ket{0\;0\;1\;0} \,.
\stoffeld

Now we consider the second detection event at position $\vec{r}_2$. The corresponding detection operator $\hat{D}({\bf r}_2)$ of this process can be written as

\begin{align}
\hat{D}({\bf r}_2) = \ket{0\;0\;0\;0}\bra{0\;0\;1\;0} + e^{i\phi({\bf r}_2)} \ket{0\;0\;0\;0}\bra{0\;0\;0\;1}\;,
\end{align}
leading to the final state 

\staffeld
\label{varphi5'}
\ket{\varphi}_{3}= \hat{D}({\bf r}_2)\,\ket{\varphi}_{2} = (e^{i\phi(\vec{r}_2)} + e^{i\phi(\vec{r}_1)}) \ket{0\;0\;0\;0} 
\stoffeld
after the detection of the second photon.

The second order correlation function can be easily calculated from the state $\ket{\varphi}_{3}$. One obtains 

\begin{align}
\label{G2calcphot}
G^{(2)}({\bf r}_1,{\bf r}_2) = | \ket{\varphi}_{3}({\bf r}_1,{\bf r}_2) |^2 \sim 1 + cos[\phi({\bf r}_2) - \phi({\bf r}_1)]
\end{align}
which shows the same modulation and thus the same non-classical behavior as Eq.~(\ref{G2calc}). Again,  introducing the visibility ${\cal V}\leq1$ incorporating experimental insufficiencies, we obtain
\begin{align}
\label{G2calcphotV}
G^{(2)}({\bf r}_1,{\bf r}_2) \sim 1 + {\cal V} \cdot cos[\phi({\bf r}_2) - \phi({\bf r}_1)]
\end{align}
in accordance with Eq.~(\ref{G2V}).

This analysis shows that although the state $\ket{\varphi}_{1}$ introduced in Eq.~(\ref{varphi3'}) is fully separable, entanglement among the spontaneously emitted photons is created by placing two detectors in the far field and by assuming that each detector measures exactly one photon. Due to this post-selective constraint we considerably reduce the photonic mode space and extract via mode selection out of the infinity of modes the maximally entangled quantum paths given in Eq.~(\ref{varphi1'}).\\

We emphasize that the photonic mode description and in particular the appearance of an entangled photon state is independent of the photon source chosen. Throughout the preceeding analysis we only considered two indistinguishable photons propagating in free space from the source towards the detectors (cf. Eqs.~(\ref{varphi3'})-(\ref{G2calcphot})), whereas the specific source of the photons was not important.\\

\begin{figure}[h!]
\centering
\includegraphics[width=0.20 \textwidth, bb=0 0 467 293]{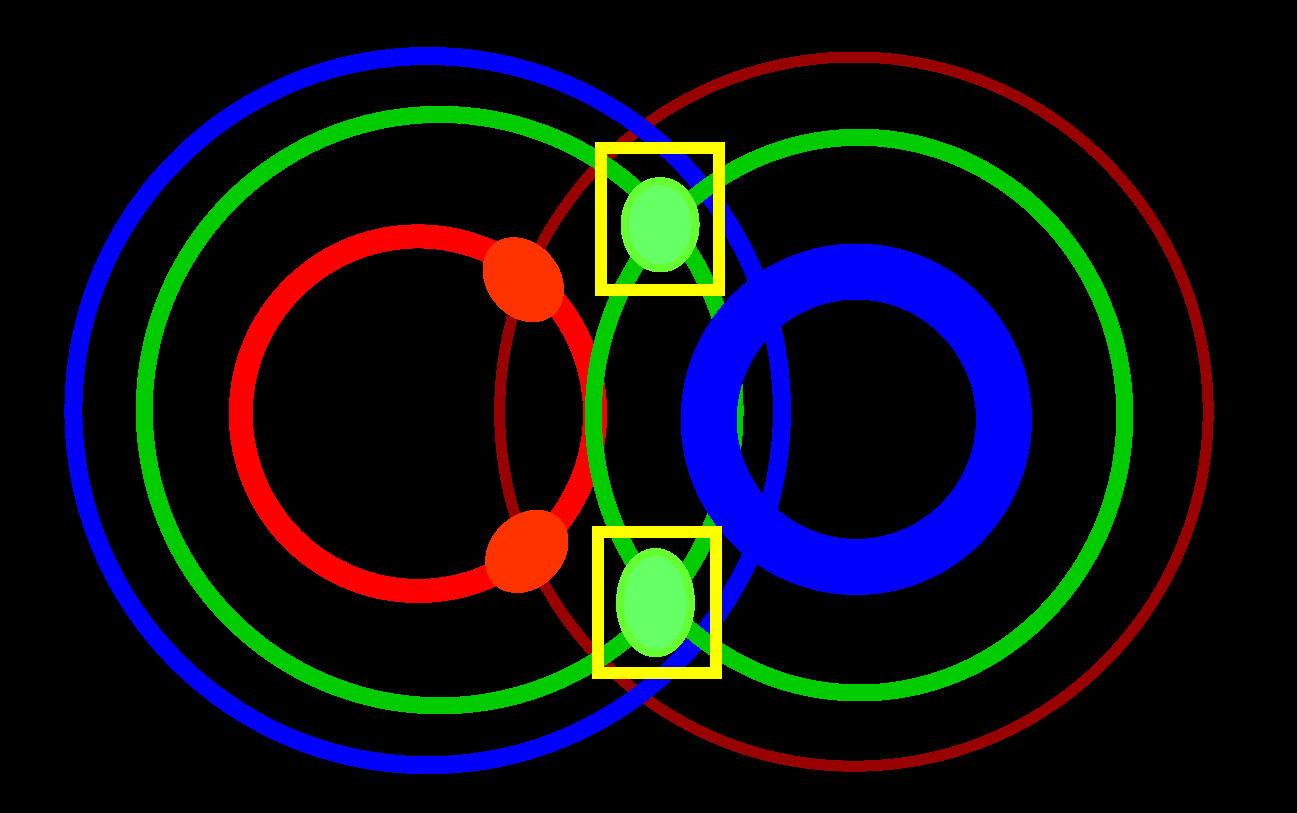}
\caption{\label{SPDCtypeII} Schematic illustration of several cone pairs generated by SPDC type II. In order to obtain polarization entangled photon pairs - corresponding to the state $\ket{H\;V} + \ket{V\;H}$ - one has to select the photon pairs emitted along the intersections of the two frequency degenerate cones as indicated by the two (yellow) rectangles.}
\end{figure}

We also note that the same kind of two-photon entanglement appears in the case of spontaneous parametric down conversion (SPDC) \cite{Zeilinger95,shihalley88}: also in this case entanglement among two photons is created via selection from an infinite set of separable two-photon quantum states. For example, in case of SPDC type II the entangled state $\ket{H\;V} + \ket{V\;H}$ is formed via the selection of two particular spatial modes, namely the modes defined by the intersection of two cones consisting of a horizontal and a vertical polarized photon (abbreviated by $H$ and $V$, respectively, cf.~figure \ref{SPDCtypeII}) \cite{Zeilinger95}. A similar mechanism leads to the entangled state emanating from SPDC type I. Here the entangled state is obtained out of a separable state by selecting certain modes and the post-selective constraint of measuring two photons at two different detectors \cite{shihalley88}.

\section{Conclusion}
\label{conclusion}

In conclusion, we investigated the correlations among indistinguishable photons spontaneously emitted by two initially  uncorrelated atoms and recorded by two detectors in the far field. If each detector registers exactly one photon it was demonstrated that by use of the second order correlation function - and without referring to auxiliary degrees of freedom as polarization - position dependent Bell-type inequalities can be maximally violated. This unambiguously displays the entanglement among the emitted photons. The entanglement was found to result from a selection of entangled photonic quantum paths out of an infinite number of possibly occupied modes due to the process of detection. The mechanism was shown to be independent of the particular photon source used and is in close analogy to the creation of entanglement among photons generated by SPDC. We note that our entanglement scheme is experimentally feasible and can be implemented with current technology, e.g., by use of quantum dots \cite{Boyd}, neutral atoms \cite{Grangier}, ions \cite{Monroe,Eschner} or molecules \cite{Ahtee}.

\section{Ackowledgements}

R.W.~gratefully acknowledges financial support by the Mayer-Foundation Erlangen and by the Elite Network of Bavaria. C.T.~thanks the Staedler foundation for funding. This work was supported by the Deutsche Forschungsgemeinschaft.

\end{document}